# Two-Dimensional Phase Transitions in Classical Systems: 60 Years after the Hohenberg-Mermin-Wagner Theorem

Ruijian Zhu(朱睿健)[1,2], Yanting Wang(王延颋)[1,2,3†]

[1] *Institute of Theoretical Physics, Chinese Academy of Sciences, 55 East Zhongguancun Road, Beijing, 100190, China*
[2] *School of Physical Sciences, University of Chinese Academy of Sciences, 19A Yuquan Road, Beijing, 100049, China*
[3] *State Key Laboratory of Surface Physics and Department of Physics, Fudan University, Shanghai, 200433, China*

In 1966, Hohenberg, Mermin and Wagner proved that long-wavelength fluctuations destabilize the long-range order of continuous symmetry in two-dimensional (2D) systems. Later in the 1970s, Berezinskii, Kosterlitz and Thouless developed the BKT theory describing an unconventional phase transition between quasi-long-range and short-range order in 2D systems driven by the binding-unbinding of topological defects, which has become a fundamental topic in statistical mechanics, condensed matter physics, and soft matter physics. One of the most important applications of the BKT theory is the melting of 2D crystals, whose mechanisms are not yet fully understood. Recently, this topic has been extended to the area of active matter, where the non-equilibrium nature leads to novel phenomena that deviate from the Hohenberg-Mermin-Wagner theorem. In this review, we first focus on the recent theoretical and computational progress in the 2D melting problem in passive systems, and then summarize the inspiring results obtained from non-equilibrium systems. The review closes with comments on several promising directions for predicting 2D melting scenarios and for understanding the non-equilibrium nature in 2D active matter systems.

**Keywords: Hohenberg-Mermin-Wagner theorem, BKT phase transition, 2D melting, active matter**

**PACS:** 64.70.dm, 05.10.Cc, 82.20.Wt, 05.65.+b

## 1. Introduction

Over 90 years ago, Peierls and Landau realized that non-negligible long-wavelength thermal fluctuations prevent the breaking of continuous symmetry in two dimensions (2D)[1, 2]. This conjecture was rigorously proven by Hohenberg, Mermin and Wagner in 1966 and is now known as the Hohenberg-Mermin-Wagner (HMW) theorem[3-5]. The HMW theorem does not exclude the existence of quasi-long-range order in 2D systems proposed by Berezinskii, Kosterlitz, and Thouless in the early 1970s[6-8]. They pioneered the concept of topological defects in the 2D XY model and revealed an unconventional type of phase transition from quasi-long-range order to short-range order, characterized by the binding and unbinding of topological defects. This is usually referred to as the BKT theory, and Kosterlitz and Thouless were awarded the Nobel

† Corresponding author. E-mail: wangyt@itp.ac.cn

Prize in 2016 for their “theoretical discoveries of topological phase transitions”[9, 10]. One of the most important applications of the BKT theory is the melting of 2D crystals. Although long-range translational order cannot be stabilized in 2D, there can be long-range bond-orientational order along with quasi-long-range translational order[8]. In the late 1970s, based on interaction between topological defects in 2D crystals, Halperin, Nelson, and Young derived a renormalization group theory[11-13], known as the KTHNY theory, in which the melting passes through an intermediate phase called hexatic[11] featured by quasi-long-range bond-orientational order and short-range translational order, which has no analogues in 3D. The KTHNY theory predicts that the transitions from crystal to hexatic and from hexatic to liquid both belong to the BKT type. Chui provided another melting scenario by considering the grain boundaries[14], which leads to a simple first-order phase transition from crystal to liquid. This melting scenario is also possible when the two phase-transition points of the KTHNY theory coincide[15].

With the development of computational techniques, molecular simulations have made great contributions to this area. By simulating hard disk systems, Bernard and Krauth revealed the third melting scenario, known as the hard-disk-like behavior[16], where the crystal melts first into hexatic via a BKT transition and then into liquid through a first-order transition, which is beyond the scope of the analytical theory. Later works focused on the influence of various monomer properties, such as shape, interaction, and deformation, on phase behaviors. Nowadays, all three melting scenarios have already been observed in a variety of systems, ranging from quantum systems[9, 17-30], confined water[31, 32], and covalent systems[33], to colloidal systems[34-41].

Active matter with the ability to convert internal energy into directional motion[42-49], as a type of out-of-equilibrium many-body system, has recently become one of the hottest topics in soft matter and statistical mechanics. The pioneering work of the “flying XY model” (the Vicsek model[50]) suggested a seeming violation of the HMW theorem[3-5] due to its non-equilibrium nature. Recent works have developed different kinds of non-equilibrium XY models[51-59] and 2D crystals[60-69] composed of active components, in which novel phenomena emerge.

Several high-quality review articles have already addressed the theoretical developments of the 1970s and 1980s, including the landmark review by Strandburg in 1988[70], a personal perspective by Kosterlitz in 2016[10], and a focused discussion of the HMW theorem by Halperin in 2019[71]. These reviews primarily concentrate on early theoretical derivations before shifting to quantum systems; accordingly, we only briefly cover these theoretical foundations in this review. While a 2023 review[72] incorporated some results from computer simulations of interacting particles, it neither systematically summarized the extensive results accumulated to date nor provided a general framework linking monomer properties to melting scenarios. Furthermore, to our knowledge, no review articles have yet focused on the recent notable findings in non-equilibrium 2D melting.

This review focuses on the progress in understanding the phase behaviors of classical 2D systems in the 60 years since the HMW theorem. The following sections are organized as follows: In Sec. 2, we briefly introduce the fundamental theories of 2D phase transitions; In Sec. 3, we review molecular simulations of 2D melting in passive matter systems, concentrating on the phenomenological rules summarized from exhaustive studies, and we then turn to the recent developments extending previous results to active matter systems in Sec. 4. The article is concluded with a brief summary and a discussion of potential directions for future work.

## 2. Theoretical Foundations of 2D Phase Transitions

The central idea underlying 2D phase transitions is the topological defect. For a given 2D continuous field $\phi(\vec{x})$, a topological defect is defined as a point satisfying the circulation condition $\oint \nabla\phi \cdot d\vec{s} = 2\pi q$, where $q$ is a non-zero integer[8]. It is obvious that this integration should be zero for a smooth $\phi$-field, so a topological defect is mathematically a singularity. Based on the Hamiltonian describing the effective interaction between topological defects, a renormalization group theory is constructed to investigate the phase transition point and the critical behavior[8, 11, 12, 73]. Below we will review the BKT theory for the XY model and the KTHNY theory for 2D melting of crystals, after which we briefly comment on the recent developments in analytical theories.

### 2.1. BKT transition in the XY model

The XY model has a unit vector restricted to the 2D plane on each lattice site, described by a pointing direction $\theta \in [0, 2\pi)$. An alignment interaction is applied between neighboring spins with the corresponding Hamiltonian written as $H = -J\sum_{<ij>}\cos(\theta_j - \theta_i)$, where the summation runs over all neighboring pairs. A topological defect is defined regarding the spin direction, i.e., $\oint (\nabla\theta)\cdot \mathrm{d}\vec{s} = 2\pi q$. Figs. 1(a)-(d) show several typical configurations of topological defects. It is worth mentioning that the point-like topological defects are unique for 2D systems, but ill-defined in 3D because a +1 defect can be continuously transformed into a -1 defect, despite the fact that a line or plane defect can still be defined in 3D.

Considering the continuous field of spin direction $\theta$, the Hamiltonian can be decomposed into a Gaussian spin-wave term $\beta H = \frac{K}{2}\int d^2x(\nabla\theta)^2$ and an effective Hamiltonian of the interaction between topological defects $\beta H = \sum_i \beta E_{c,i} - 4\pi^2 K \sum_{i<j} q_i q_j C(\vec{r}_i - \vec{r}_j)$, where $K = \beta J$, $C(\vec{r}) = \ln(|\vec{r}|)/2\pi$ and $E_{c,i}$ is the self-energy of each topological defect (core energy)[8, 70, 74], sharing the same formula with the 2D Coulomb gas. Consequently, the partition function can be separated into a trivial spin-wave term and a non-trivial 2D Coulomb gas term. Topological defects gather into dipoles or larger clusters to reduce the energy cost at lower temperatures, resembling the behavior of charges in the dielectric; and unbind at higher temperatures benefiting from the larger entropy, similar to the electrons in the metal. This "dielectric-to-metal" transition is the central picture of the BKT theory[74]. The renormalization group (RG) analysis of the 2D Coulomb gas suggests that, at lower temperatures, the recursion terminates at a non-universal point between 0 and $\frac{\pi}{2}$ on the $K^{-1}$-axis, while at higher temperatures, the recursion diverges, i.e., $K$ terminates at zero[8, 74, 75]. This behavior is unconventional since it seems that every $K^{-1}$ between 0 and $\frac{\pi}{2}$ is a critical point. Indeed, the order parameter $S \equiv |\sum_j e^{i\theta_j}|$ always takes a zero value in the thermodynamic limit but exhibits a quasi-long-range correlation without a characteristic length scale below the phase transition point, resembling the critical-point-like behavior. The renormalization value of $K$, $K_{\text{eff}}$, referred to as the stiffness, can be directly measured in the simulation of the XY model in terms of $K_{\text{eff}} = \beta\gamma = \frac{\beta}{L^2}[<\sum_{<ij>_x}\cos(\theta_i - \theta_j)> - K < (\sum_{<ij>_x}\sin(\theta_i - \theta_j))^2 >]$, where the summation is performed over all links in one direction and $\gamma$ is called the helicity modulus[76, 77]. In experiments on superfluid films, the stiffness is related to the superfluid density $\rho_s = m|\psi|^2$ by $K_{\text{eff}} = \frac{\beta\hbar^2\rho_s}{m^2}$ [76, 78], which indeed exhibits a universal jump of $\frac{2}{\pi}$ at the transition[78-80].

When the system temperature approaches the phase transition point, the free energy has only an essential singularity, which results in no divergence of any order of its derivatives, manifesting that the BKT phase transition is an infinite-order phase transition according to Ehrenfest's definition. The numerical simulation on the XY model[81] has indicated that the heat capacity shows a peak at a temperature higher than the phase transition point, resulting from the unbinding of a large number of topological defects, while the phase transition point theoretically corresponds to the temperature at which only a single pair of topological defects unbinds.

Numerical determination of the critical point and the critical exponents usually suffers from the finite-size effect and the critical slowing-down. The finite-size scaling allows an accurate estimation based on the simulations at several relatively small systems. Weber and Minnhagen[77] provided an ansatz of $\gamma(L) = \gamma_\infty(1 + \frac{1}{2}\frac{1}{\ln L + \text{C}})$ where $L$ is the number of sites on one dimension, $\gamma$ is the helicity modulus and $\gamma_\infty = \lim_{L\to\infty} \gamma(L)$. The simulation results at various temperatures and system sizes suggested a fitted critical point at $K^{-1} = 0.887$. Hasenbusch[82] later found that the fitting parameters in the scaling of $\gamma$ depend strongly on the system size, but exhibit a robust result of $K^{-1} = 0.8929$ considering the scaling of the second-moment correlation length of spins, in agreement with the Monte Carlo RG method[83] employing the exact duality between the XY model and a solid-on-solid model[84] as well as the large-scale simulation[85] using the probability-changing cluster algorithm[86]. Alternatively, the short-time dynamics approach employs relatively short simulation trajectories combined with the finite-time scaling, successfully avoiding the extremely slow relaxation. It is important that there are different dynamic scaling laws for different initial states[87, 88]. Starting with a perfectly aligned state, Zheng et al.[89] determined the critical point as $K^{-1} = 0.8942$, and extracted the two critical exponents regarding correlation length and susceptibility, approximately coinciding with the analytical theory[73]. Recently, Zuo et al.[90] derived a series of scaling theories for BKT phase transitions based on the distance from the critical point instead of correlation length. Although some of their conclusions are verified in a quantum model, they, especially the new finite-size scaling, require further numerical verification.

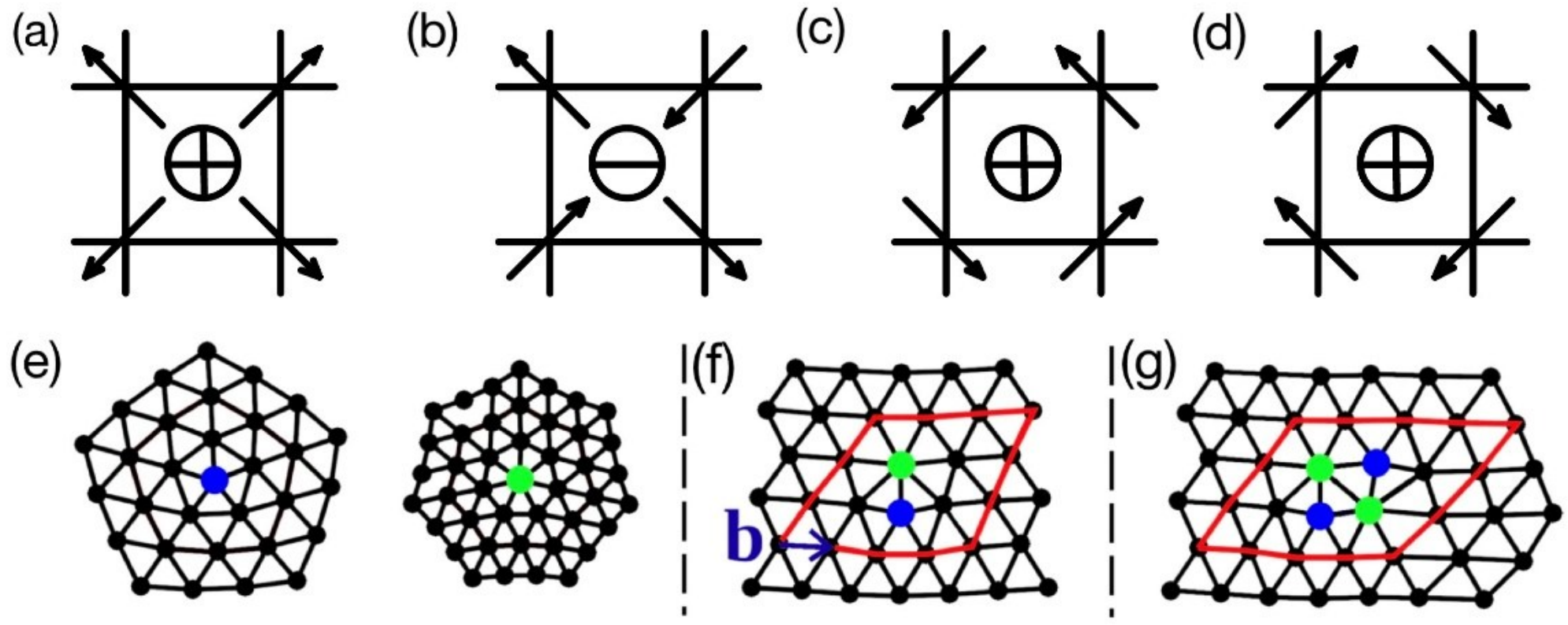

Fig. 1 Schematics of topological defects in spin systems ((a)-(d)) and 2D crystals ((e)-(g)). (a) A topological defect with a positive charge. (b) A topological defect with a negative charge. (c) An anti-clockwise vortex. (d) A clockwise vortex. It should be noted that both vortices are positive charges. (e) The blue and green dots mark the points with 5 and 7 neighbors, respectively, defined as disclinations with positive and negative charges. (f) Two disclinations with opposite signs form a dislocation characterized by a Burgers vector $\vec{b}$. (g) A dislocation pair is composed of two dislocations with opposite Burgers vectors. Panels (e)-(g) adapted from ref. [91].

## 2.2. Melting of 2D crystals

The free energy of a crystal is invariant under a continuous transformation $\vec{u}(\vec{r}) \to \vec{u}(\vec{r}) + \vec{u}_0 + \theta_0(\hat{z} \times \vec{r})$, representing the symmetry under a translation of $\vec{u}_0$ and a rotation of $\theta_0$ to the original displacement field $\vec{u}(\vec{r})$ [92]. The rotational symmetry can also be reflected by considering the field of bond-orientation between neighboring particles, defined as $\theta(\vec{r}) = -\frac{1}{2}\hat{z}\cdot\nabla\times\vec{u}$ [74]. Although these two symmetries always vary simultaneously in 3D, it is possible that the rotational symmetry breaks before the translational one in 2D.

The topological defect can be defined in terms of either translational symmetry or rotational symmetry. A more direct way is to consider the number of neighboring particles. Taking the triangular lattice as an example, a disclination is defined as a particle with 5 or 7 neighbors, which can be viewed as a positive or a negative charge, respectively. A dislocation is composed of two disclinations with opposite signs, characterized by a non-zero Burgers vector, while two dislocations with opposite Burgers vectors combine into a dislocation pair, which can be regarded as a quadrupole of disclinations. These topological defects in real space are visualized in Figs. 1(e)-(g). The interaction between two disclinations shares the same formula as that of the 2D Coulomb gas. The dislocations can be viewed as a "vector Coulomb gas", where the Burgers vector plays a role as charge and leads to a Hamiltonian of $\beta H = -\frac{\overline{K}}{2}\sum_{i,j} b_i^\alpha b_j^\beta C_{\alpha\beta}(\vec{r_i}-\vec{r_j})$ where $C_{\alpha\beta}(\vec{r}) = \frac{1}{2\pi}[\delta_{\alpha\beta}\ln(\frac{|\vec{r}|}{a}) - \frac{r_\alpha r_\beta}{r^2}]$ and $\overline{K}a^2 = \frac{2\mu(\mu+\lambda)}{2\mu+\lambda}$ is a combination of the two Lamé constants $\mu$ and $\lambda$ [8, 11, 12, 74].

It is possible that the crystal melts by following the KTHNY melting scenario[8, 11-13] including two BKT-type phase transitions. Dislocation pairs are induced by unavoidable long-wavelength thermal fluctuations in 2D crystals at finite temperatures, resulting in a quasi-long-range translational order and a long-range bond-orientational order[8]. With increasing temperature, the translational order disappears when a dislocation pair unbinds into two free dislocations, and the crystal simultaneously melts into a hexatic phase characterized by the quasi-long-range bond-orientational order[11, 12]. Further heating leads to the unbinding of dislocations into free disclinations, which drives the system into a liquid state with both short-range translational order and bond-orientational order.

Another melting scenario is that the crystal directly melts into liquid discontinuously due to the simultaneous formation of a line of free dislocations, i.e., a grain boundary. This mechanism was first proposed by Chui in 1982[14], where he drew a strong conclusion that 2D melting is always a first-order transition. However, his central assumption is violated when there are only a small number of topological defects, so it is expected that the melting scenario should depend on whether it is "easy" to create a topological defect, which can be quantified by the core energy $E_c$. Chui's theory provides a boundary of $E_c = 2.84kT$, close to 2.7 $kT$ obtained by the later simulations[70, 93, 94] on the Laplacian roughening model mimicking the Coulomb gas[92].
It is somewhat surprising that the RG analysis of the square lattice has not yet been fully established. The central difficulty lies in the fact that the square lattice has one more elastic modulus than the triangular lattice, resulting in more complex coupling and RG flow. Not until last year did a preprint[95] advance the understanding of this problem, in which the authors theoretically verified the tetratic phase[96] (defined similarly to the hexatic, but with 4-fold symmetry), which has already been observed in simulations[96-102] and experiments[103-105]. More importantly, they discovered that the decay exponent of the translational correlation function in the square lattice should be smaller than 1/4, rather than 1/3 in the triangular lattice[8, 74, 106]. The upper bound of the decay exponent of bond-orientational correlation function in the tetratic phase takes the value of 1/4, the same as the one in the hexatic phase.
The hard disk system is one of the simplest models in statistical mechanics. However, simulations on the melting scenario of hard disks[106-114] did not come to an agreement until 2011. In 2008, colloidal experiments performed by Han et al.[34] raised the possibility of a first-order hexatic-to-liquid phase transition, out of the expectations of all previous analytical theories. This was confirmed by Bernard and Krauth in 2011 via large-scale Monte Carlo simulations[16], revealing the third melting scenario, where crystal melts into hexatic continuously and then into liquid discontinuously. This scenario is usually referred to as hard-disk-like behavior.
How a specific 2D system melts still remains elusive. Early analytical theories[14, 70, 92, 93] relied on continuum descriptions, where the core energy is obviously a collective quantity, which fails to connect to microscopic details. Recently, some attempts combining the calculation of different elastic moduli with molecular details[111, 115, 116] and the HNY RG analysis[11-13] are expected to build up the connection. However, even for a simple Gaussian core model, the analytical theory significantly diverges from simulation results at high densities[115]. Some fundamental difficulties have been raised in an early review[70]. More difficulties include: (1) The picture of the KTHNY theory is too simple in some cases to include complex structures of topological defects, which will be described in Sec. 3; (2) There is no lattice structure for the hexatic state; (3) The interaction strength of disclinations in the hexatic phase is described by the Frank modulus, which has not been connected to molecular details; (4) An RG formula for complex lattices like Kagome would be extremely complicated. Fortunately, a huge number of simulations provide insightful results, making qualitative predictions of melting scenarios based on molecular details possible. We will summarize these phenomenological rules in the following section.

## 3. Molecular Simulations of 2D Melting in Passive Systems

The computer power nowadays can largely overcome long relaxation times and the finite-size effects, making direct simulations of various systems possible. Molecular dynamics simulations and different kinds of Monte Carlo simulation methods have been employed, for which we refer to several classical textbooks[117-119]. Below we will first discuss how to identify the phase transition and different phases, and then we will summarize the results coming from simple models with only interaction or shape.

Combining these two factors together makes the model more realistic, and indeed reveals more interesting results. The consideration of substrate and impurities is instructive for experiments. Finally, we will talk about the numerical verification of analytical theories.

### 3.1. Numerical methods for determining phases and phase transitions

Multiple methods can be used to identify a first-order phase transition, including the Mayer-Wood loop[120], two-phase coexistence, and the hysteresis in heating and cooling. Besides heating, an equivalent way to drive phase transition is increasing the system volume, which is a convenient way for rigorously determining the phase transition order. The choice of ensemble is critical, dictating a stark contrast in phase behavior: the *NVT* ensemble permits spatial coexistence, manifesting as a characteristic loop in the pressure-volume (*P*-*V*) curve. In contrast, the *NPT* ensemble fundamentally reshapes this into temporal coexistence, which manifests as a discontinuous jump in the *P*-*V* curve. This difference arises from the metastability of spatial coexistence in a finite system due to interfacial energy. Strictly speaking, for a first-order phase transition, the loop area (which can only be calculated in an *NVT* simulation) must satisfy a scaling relation with the inverse-square-root of the particle number[120]. A continuous transition in 2D melting is always a BKT one.

Binder cumulants are widely used to determine the phase transition point and order in active matter systems, but whether they are suitable for characterizing 2D melting is not fully established[111, 113, 114]. A more concrete discussion on Binder cumulants can be found in a recent review[72].

The determination of phases relies on two order parameters: the translational order parameter $S \equiv \sum_j s_i = \sum_j e^{i\vec{G}\cdot\vec{r}_j}$ reflecting the translational symmetry and the bond-orientational order parameter $\Psi \equiv \sum_j \psi_j = \sum_j \frac{1}{n}\sum_{k=1}^{n} e^{in\theta_{jk}}$ quantifying the rotational symmetry, where $n$ is the number of nearest neighbors determined by the Voronoi tessellation and $\theta_{jk}$ is the angle between a neighbor $k$ and the central particle $j$ with respect to the $x$-axis. The order parameter takes a non-zero value for long-range order and zero for short-range order. Theoretically, it should be zero for a quasi-long-range order, but it practically remains finite in simulations since the thermodynamic limit can never be reached. Consequently, the order parameter serves to identify the discontinuous phase transition from quasi-long-range to short-range order, whereas the correlation function is essential for characterizing continuous transitions. While the bond-orientational correlation function is straightforward to calculate, the reliable calculation of the translational one has only been developed in recent years. Early works used the reciprocal vector $\vec{G}$ of the perfect lattice at a given density, which was later found to incorrectly yield an exponentially decaying translational correlation even in the crystal state[16, 121]. An iteration procedure proposed by Han et al.[34] and later modified by Bernard and Krauth[16] scans the reciprocal vector plane to identify the vector $\vec{G}$ corresponding to the peak value of the translational order parameter *S*. The scanning result of a hard-disk system is shown in Fig. 2 (a), exhibiting a clear shift of $\vec{G}$ with respect to the perfect lattice, which is believed to arise from the existence of vacancies and other defects[122]. A comparison of the computational results using the reciprocal vector determined by scanning versus that of the perfect lattice is shown in Fig. 2 (b), where the former suggests an algebraic decay as expected, while the latter indicates an exponential one. Bernard and Krauth have also proposed an alternative expression for the translational correlation function[16], $g(r)=\frac{1}{N}\sum\delta(r-(x_j-x_k))\delta(r\tan\theta-(y_j-y_k))$, where $\theta$ is the crystalline axis. This formula reflects the invariance under translation along basis vectors of a crystal. The remaining problem is the determination of $\theta$. For hard disks, it can be determined by the argument of $\Psi$, which however is incorrect for the Lennard-Jones (L-J) system[123], as depicted in Fig. 2 (c) and (d). Li and Ciamarra[121] suggested that the crystalline axis direction corresponds to the peak closest to zero degree obtained by scanning $\theta$ at a fixed *r* in $g(r)$. This procedure is illustrated in Fig. 2(e), and the results in Fig. 2 (f) are consistent with the theoretical prediction[121]. This method has been tested in more complex ball-stick polygon systems[124], yielding a good agreement with the analytical theory.

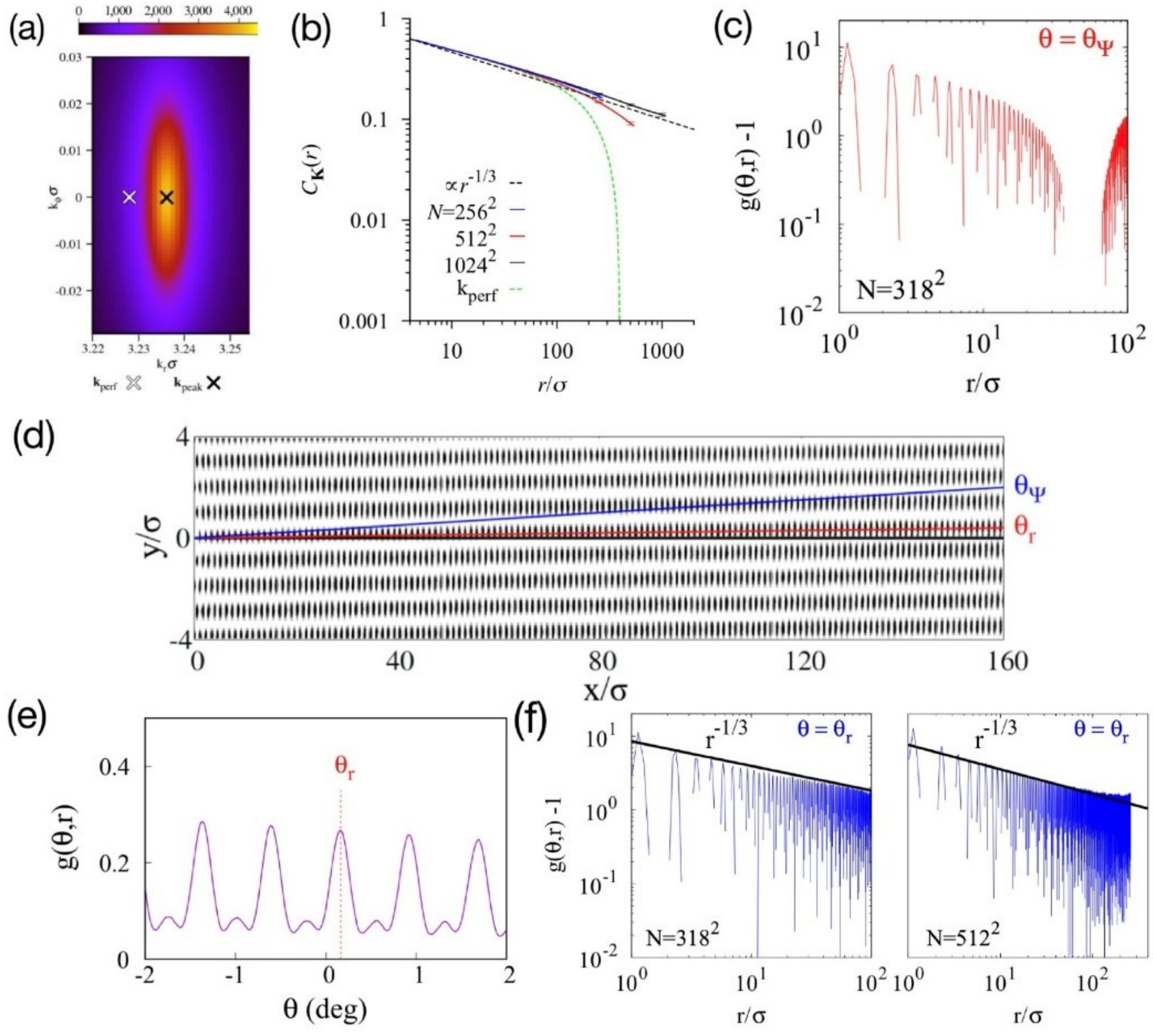

Fig. 2 Numerical procedure of the calculation of translational correlation function. (a) Translational

order parameter scanned on the reciprocal lattice vector plane. $k_{\mathrm{perf}}$ corresponds to the perfect lattice at a given density, while the maximum value is taken at $k_{\mathrm{peak}}$. (b) Comparison between the translational correlation functions calculated respectively by $k_{\mathrm{perf}}$ and $k_{\mathrm{peak}}$. (c) Translational correlation function calculated by using $\theta_{\Psi}$ as the crystalline axis shows an exponentially decaying behavior in L-J systems. (d) Comparison between $\theta_{\Psi}$ and the crystalline axis $\theta_r$. (e) Determination of $\theta_r$ by scanning the translational correlation function at a fixed distance. (f) The translational correlation function by using $\theta_r$ as the crystalline axis shows an algebraically decaying behavior. Panels (a)-(b) adapted from ref. [16]. Panels (c)-(f) adapted from ref. [121].

### 3.2. Phenomenological rules revealed by molecular simulation

Computer simulations reveal that the melting scenario is sensitive to monomer details. Below we will discuss some simulation results and summarize a set of phenomenological rules, aiming at a qualitative rather than quantitative prediction.

#### 3.2.1. Interacting point-like particles

In the 1990s, the simulation[125] on the melting scenario of a simple square-well potential system revealed the dependence on interaction parameters. Typically, an interaction potential contains an intermediate-range attraction as well as a short-range repulsion. In 2015, Kapfer and Krauth[126] performed simulations on systems composed of purely repulsive particles interacting with $U(r)=\varepsilon(\sigma/r)^n$ of various *n*. As shown in Fig. 3(a), they observed that the melting scenario changes around *n* = 6. For *n* > 6, the interaction is "hard enough", resembling the hard-disk-like behavior, while for *n* < 6, the melting obeys the KTHNY theory. The robustness of their conclusion was verified by tuning the screening length in the Yukawa potential and mapping it to a power-law repulsion, by which both KTHNY and hard-disk-like behaviors are realized.

Detailed investigations on soft-core repulsive particles revealed very complex behaviors[102, 127-135]. Several common model systems including the Hertzian potential, the harmonic potential, and the Gaussian-core model, all exhibit a reentrant phase transition[127], as shown in Fig. 3(b) and (c), in which a liquid is compressed into a crystal state, and remelts into a liquid state upon further compression. In all the above three systems, the melting scenario is hard-disk-like under low densities, while it becomes the KTHNY one at high densities. This seems not universal, since in a square-shoulder-square-well potential system, the melting scenario obeys the KTHNY theory at low densities, while it experiences a first-order solid-liquid transition at a high density. Moreover, novel crystalline structures are formed in soft-core potential systems, like Kagome lattice[136], cluster crystal[132], and quasicrystal[134, 137, 138], which make it more difficult to reach a general conclusion for the melting scenario. The determination of the melting scenario in these cases also faces technical challenges. For example, when the unit cell contains more than one particle, it remains ambiguous whether the hexatic phase should be defined with respect to the unit cell or individual particles. In the former case, it is unclear how to appropriately identify the unit cell in the hexatic phase, where particles can move freely and break the lattice structure.

The melting scenario of L-J particles, one of the simplest potentials with both repulsion and attraction, was not clarified until 2020. Benefiting from the accurate calculation of translational correlation function, Li and Ciamarra[139] scanned the phase diagram of the L-J system in the $T$ - $\rho$ plane. The phase diagram is shown in Fig. 3(d). An important result is that the coexistence region of hexatic-liquid becomes wider at lower temperatures and finally overrides the hexatic phase, leading to a solid-liquid first-order transition instead of the hard-disk-like behavior.

The universality of phase transitions tells us the interaction range (the potential decays faster or slower than $r^{-2}$) also affects the phase behavior[140, 141]. For example, the fully connected Ising model always exhibits a phase transition regardless of the spatial dimension[140]. The long-range Coulomb repulsion results in the famous Wigner crystal[142], which has truly long-range translational order and makes no guarantee to obey the KTHNY theory. A recent simulation, however, observed the hexatic phase during the melting of the Wigner crystal[143]. The melting from crystal to hexatic and from hexatic to liquid are both suggested to be weak first-order phase transitions.

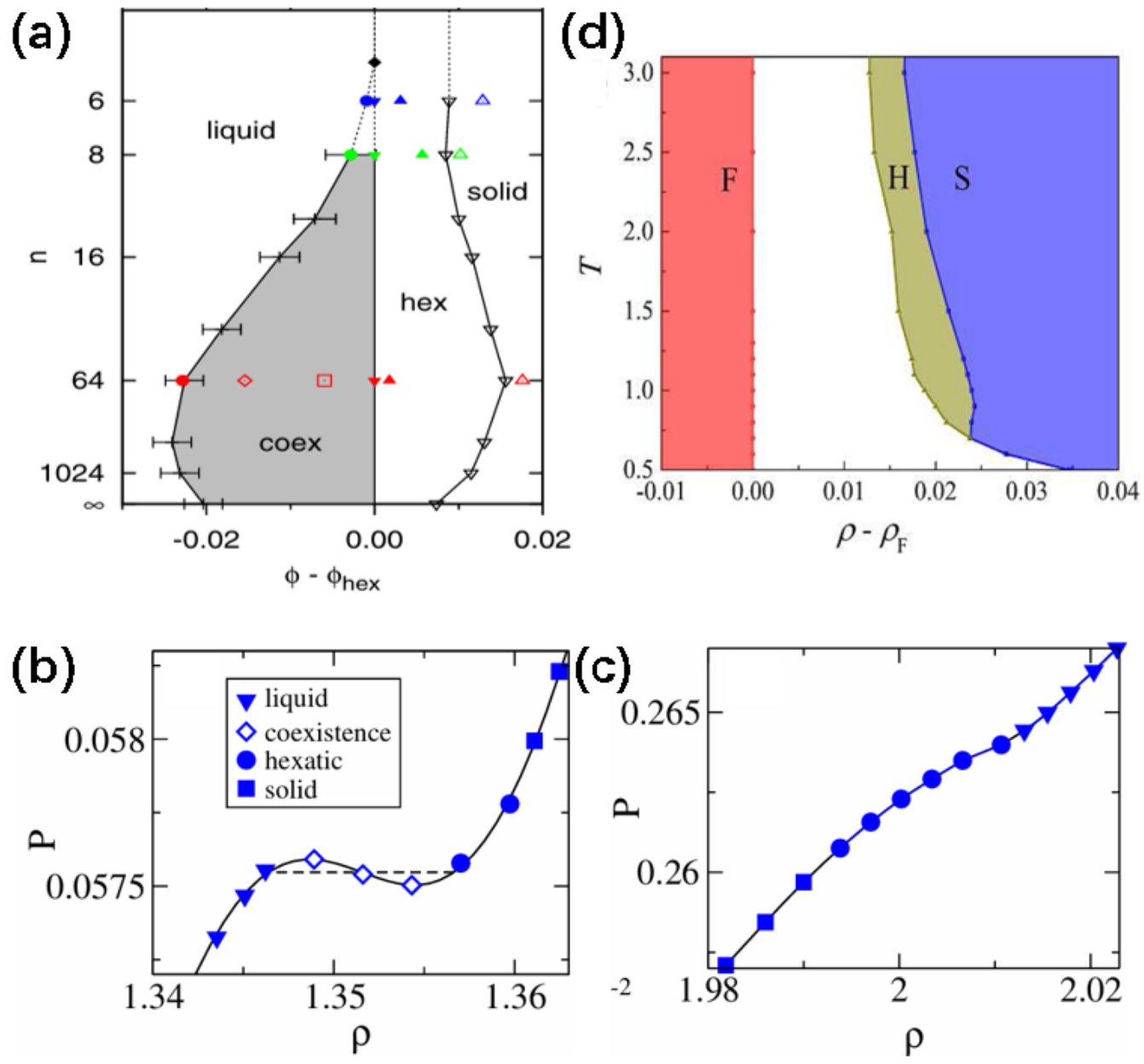

Fig. 3 Phase behaviors of several interacting point-like particle systems. (a) Phase diagram of the particle system with power-law repulsive interparticle interactions with various exponents $n$[126]. Systems with $n < 6$ exhibit a KTHNY melting scenario, while those with $n > 6$ resemble a hard-disk-like behavior. (b) and (c) are the plots of the equation of state for the Hertzian potential system,

suggesting a reentrant transition under continuous compression[127]. (d) Phase diagram of the L-J particle systems[139]. The coexistence region colored in white always exists and becomes wider at a lower temperature, which finally overrides the hexatic region.

### 3.2.2. Hard polygons

Particles with only volume repulsion form ordered structures under high density due to the larger number of accessible states, usually referred to as "entropy-induced order"[144, 145]. Therefore, different polygons offer us a good platform to understand the effect of entropy[35, 96-99, 146]. An important contribution was the observation of the tetratic phase in the simulation of hard squares[96]. In 2017, Anderson et al.[97] performed extensive Monte Carlo simulations of regular polygons from 3-gon to 14-gon. With an additional focus on the space-filling pentagon, the authors revealed that it is the orientational entropy and local symmetry in crystal and liquid phases, rather than space-filling, that determine the melting scenario. The conclusions shown in Fig. 4(a) are summarized as follows: (1) For polygons with 7 or more edges, the orientational entropy is weak, resulting in the hard-disk-like behavior; (2) For polygons with fewer than 7 edges, the KTHNY melting scenario occurs if the local symmetry in liquid and crystal states matches (triangle, square, and hexagon), while a first-order solid-liquid transition occurs otherwise (regular and space-filling pentagons).

The simulation results above are not always commensurate with the colloidal experiments for the corresponding shapes, for example, colloidal pentagons[39] suggest a frustrated glass state at high density, and colloidal squares[147] form a rhombic lattice instead of a square lattice as well as show no tetratic phase. Some of these inconsistencies have been verified to arise from the roundness of colloids[35, 98]. Irregular polygons exhibit more complicated phase behaviors, which sometimes even cannot crystallize[148, 149]. Selecting an appropriate parameter in the extensive shape space, one can obtain a continuous transformation of shape, providing more information than the discrete edge number. Recently, by symmetrically truncating a rhombus, Jiang et al.[150] simulated the melting behavior from a rhombic to a hexagon to a stick, and the melting behaviors are summarized in Fig. 4(b). The most impressive finding is that, in some cases, the system exhibits a previously unobserved two-step melting scenario, in which crystal melts into hexatic discontinuously but into liquid continuously (denoted as IV in Fig. 4(b)). It is important to further explore this melting scenario in other systems and understand whether the first-order crystal-hexatic transition is driven by grain boundaries. Here, we emphasize that the first column labeled as IV and all the columns labeled as V in Fig. 4(b) consider melting pathways that involve liquid crystalline phases, which however, should be categorized as liquid in 2D melting since their centers of mass are arranged in a totally disordered way.

Incorporating deformation, a common feature of realistic systems, leads to significant changes in the melting scenario. The Voronoi model[151-154] mimicking the cell tissue suggests a region where the hexatic is more stable than the crystal at $T$ = 0[155]. Moreover, the pressure in this model can be negative because Voronoi tessellation always fills space completely. Deeper investigations of this model typically incorporate non-equilibrium elements to enable comparison with realistic cellular systems, which we will discuss briefly in Sec. 4. In general, the phase behavior of deformable systems remains poorly understood, as these systems can be viewed as mixtures of polydisperse irregular polygons. Recent studies of polygon mixtures[156], which are still much simpler than the Voronoi model since there are only two or three types of regular polygons, indicated richer behaviors compared to monodisperse systems.

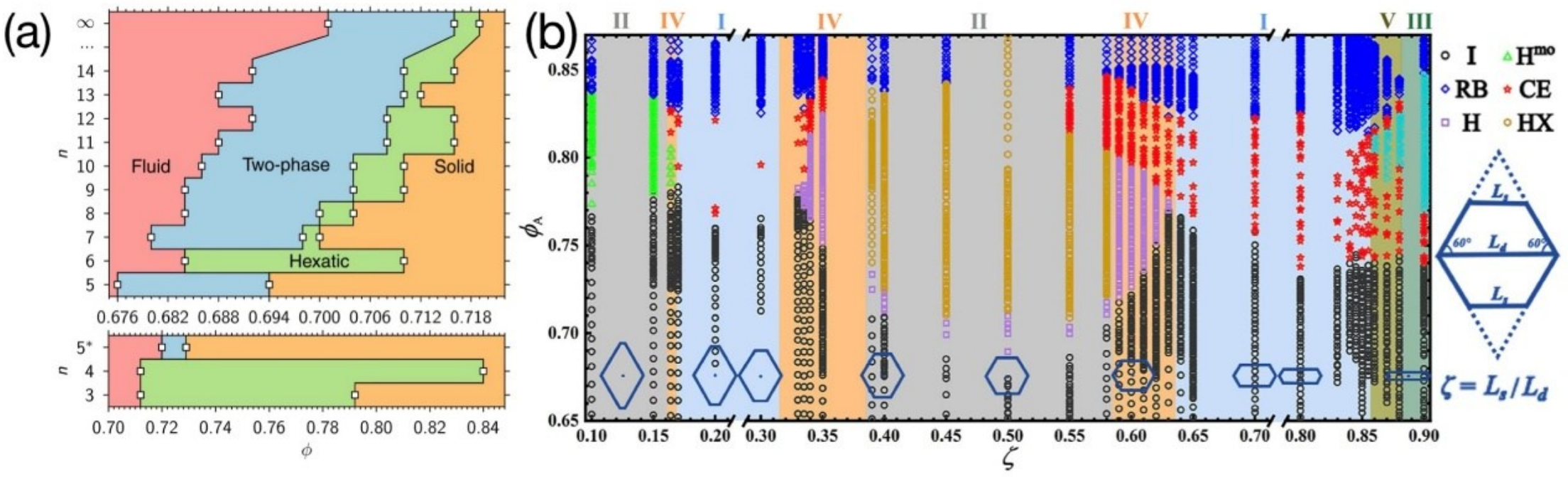

Fig. 4 Phase behavior of hard polygon systems. (a) Phase diagram of the hard regular polygon system[97]. The upper region includes all non-space-filling polygons, while the lower part shows the cases of space-filling polygons. Here, 5* denotes the space-filling pentagon. (b) Phase diagram of the truncated rhombic polygon system[157]. I to III correspond to conventional melting scenarios, and the two rightmost columns of IV correspond to the newly identified behavior. By definition, all melting scenarios involving liquid crystals ($H^{mo}$) are excluded from consideration here.

### 3.2.3. Phenomenological rules

Realistic materials, especially soft matter systems, are characterized by the balanced interplay between enthalpy and entropy[158]. As famously stated, "More is different"[159], and the simple combination of insights gained from point-like particles and hard polygons may fail to predict the behavior of systems composed of monomers with both shape and interaction properties. Therefore, designing appropriate models to capture these two key elements is essential for extending model systems toward real-world applications. Specifically, these studies are expected to provide valuable insights for 2D molecular crystals[160-163], which are stabilized by intermolecular interactions rather than covalent bonds and have become a hot topic in materials science and chemical engineering in recent years.

When entropy and enthalpy meet, they compete both spatially and thermodynamically. Li and Ciamarra[164] simulated attractive squares, pentagons, and hexagons, in which each polygon is constructed by placing L-J particles equally spaced along the perimeter, as shown at the bottom of Fig. 5(a). Their phase diagrams suggest that a first-order solid-liquid transition is universal for all cases at low temperatures, while it reproduces the melting scenario of the corresponding hard polygon at high temperatures. Zhou et al.[41] studied the phase behavior of polygons with magnetic interactions, where each polygon has a magnetic moment along the $z$-axis resulting in an anisotropic interaction in the $x$-$y$ plane. Experimental and simulation results revealed the role of spatial competition between shape and interaction, for example, the squares can form a triangular lattice at intermediate density (shown at the top of Fig. 5(a)) and the triangles are trapped into a glassy state at high density.

Zhu and Wang[124] proposed a ball-stick regular polygon model, with an L-J ball on each vertex, capturing all four elemental aspects through the intrinsic spatial competition in this model, in which the close-packing scheme of the corresponding polygon is forbidden by the L-J ball, as shown in Fig. 9 in ref. [124]. This idea is schematically illustrated in Fig. 5(a). A critical edge number $n_c = 4$ is revealed in this model, as shown in Fig. 5(b). Below this value (i.e., for the ball-stick triangle), no crystalline morphologies exist at any finite temperatures. The critical case corresponds to the ball-stick square, which forms a distorted square lattice enriched with topological defects at a concentration of approximately 50%—far exceeding the typical value at the phase transition point confirmed by several simulations[156, 165]. Moreover, the topological defect concentration in the ball-stick square system is robust even at temperatures close to zero, manifesting its intrinsically disordered nature rather than disorder driven by external thermal noise. These two polygons both melt discontinuously into the liquid phase at arbitrary pressures, which is attributed to the sufficiently high defect concentration in their crystal states. For pentagon, hexagon, and octagon, a normal crystal state can be stabilized at finite temperatures, whose phase diagrams are shown in Fig. 5(c)-(e). The hexatic phase always vanishes at low pressures but emerges for hexagons and octagons at high pressures.

Compared with the study on LJ-bead polygons, we can summarize a set of phenomenological rules governing the emergence of the hexatic phase:

(1) If the crystal state is intrinsically disordered, the hexatic phase is absent; otherwise, the rules below apply;

(2) At low temperatures/pressures, the hexatic phase is always replaced by a broad solid-liquid coexistence region;

(3) At high temperatures/pressures, the existence of the hexatic phase follows the same phenomenological rules as for the corresponding hard polygons, i.e., dominated by shape.

Next, we turn to the phase transition order. It is interesting that the ball-stick hexagon melts via a hard-disk-like scenario at high densities[124]. This is different from the hard hexagon[97] and the LJ-beads hexagon[164] following the KTHNY theory, but is consistent with the simulation result of round-corner hexagons[35], which may be attributed to the existence of the rotator crystal, a crystal phase without body-orientation. Before melting, the ball-stick hexagon experiences a solid-solid transition from the triangular lattice with ordered body-orientation to the rotator crystal phase. A similar solid-solid transition has also been observed in the round-corner hexagon system but not in the hard hexagon system. The loss of body-orientation order in the rotator crystal can significantly weaken the orientational force. Shen et al. studied the rotator crystal in hard polygon systems with $n$ = 3~12 and revealed three cases[166]: there are no rotator crystals when the local symmetry group of the lattice is a subgroup of that of the particles ($n$ = 3, 4, 6, 12); there is a discrete rotator crystal phase with several preferred body-orientations when these two symmetry groups share some common elements ($n$ = 8, 9, 10); there is a continuous rotator crystal with uniform body-orientation distribution otherwise. However, the rotator crystal is suggested to be universal in systems where the interaction range can exceed the nearest neighbors[167], and is indeed observed in the magnetic polygon system[167] and the ball-stick polygon system with $n$ = 5~13[124, 168]. Most previous work on 2D melting does not distinguish different crystal phases, so the role of the rotator crystal in the melting scenario remains largely unexplored. Consequently, we suggest that more attention should be paid to this mesophase. It is very likely that the local symmetry determines the appearance of the hexatic phase, while the orientational force, which depends on whether there is a rotator crystal phase, determines the phase transition order.

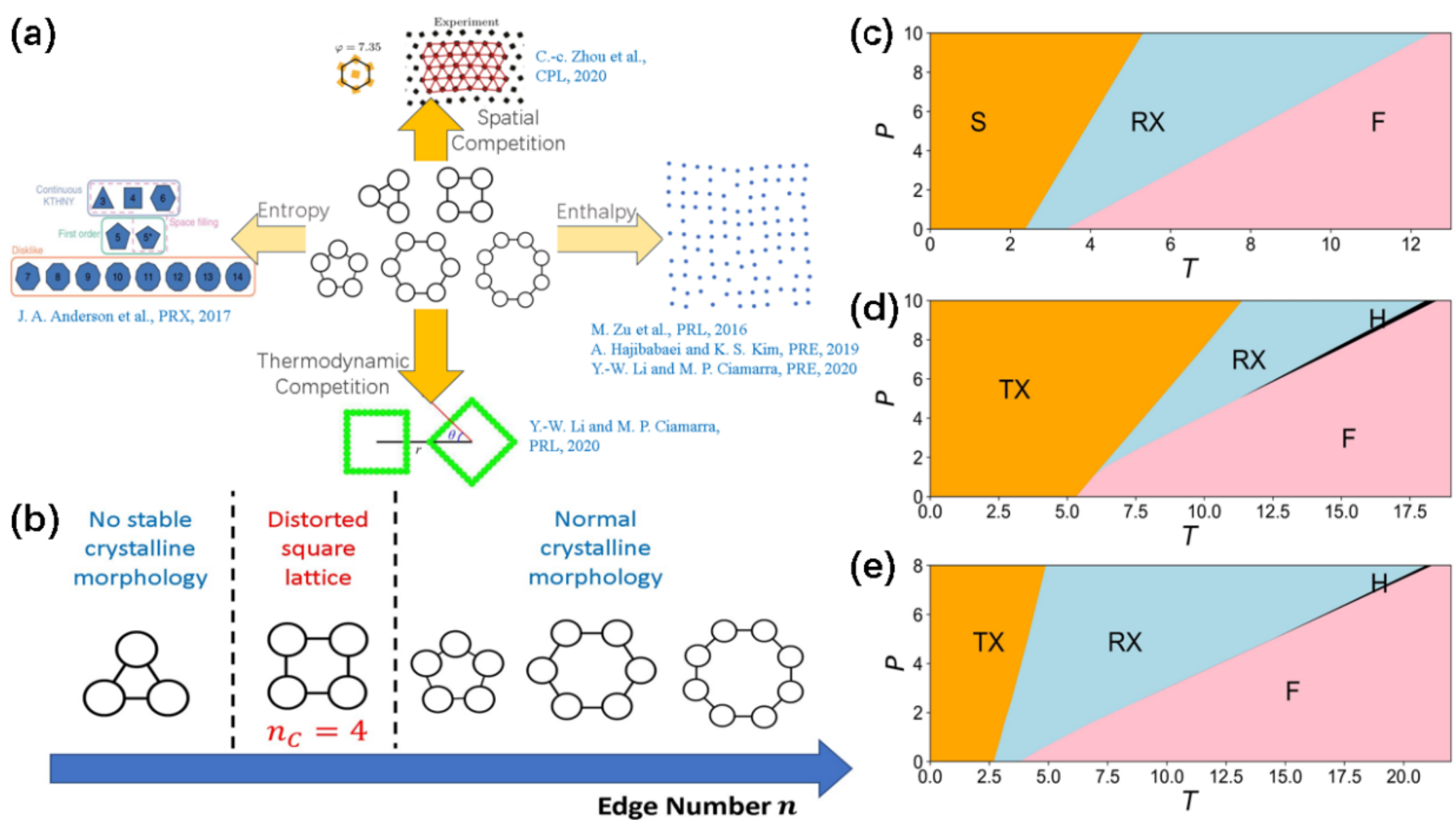

Fig. 5 Phase behaviors of the ball-stick polygon systems. (a) Ball-stick polygon model captures four important physical elements in terms of entropy and enthalpy as well as the spatial and thermodynamic competitions between them. Subfigures adapted from refs. [41, 97, 164]. (b) Summary of the ground-state morphologies of several ball-stick regular polygons[124]. (c)-(e) Phase diagrams of the ball-stick pentagon (c), hexagon (d), and octagon (e)[124]. S represents the striped phase, TX the triangular lattice with unique body-orientation, RX the rotator crystal phase, H the hexatic phase, and F the fluid phase. It should be noticed that a coexistence region cannot appear in a $P$ - $T$ phase diagram, but here we never observe a KTHNY melting scenario, i.e., all the solid-liquid and hexatic-liquid transitions are discontinuous.

### 3.3. Effect of substrate and impurity

2D materials are usually supported by a substrate. Shortly after the establishment of the original KTHNY theory, the effect of substrate was theoretically studied by introducing in the model an additional term describing the alignment between the axes of crystal and the substrate[12, 13, 70]. This coupling term modifies both the critical exponent of the crystal-hexatic transition and the decay exponent of the positional order in the crystal phase[12, 13, 70]. More interestingly, the bond-orientational order remains long-ranged above the melting temperature in such systems[169]. Furthermore, when a triangular lattice crystal is placed on a square substrate, the BKT transition is replaced by an Ising-like phase transition[12]. It is expected that a well-designed substrate can significantly change the melting scenario. Several experiments have suggested that the elasticity of the substrate boundary, surface tension with a specific phase, and detailed structure of the substrate can influence the melting scenario as well as the spatial patterns in the coexistence region[170, 171].

Impurities caused by the substrate or intrinsic imperfections are common in experiments of 2D crystal systems. By theoretically introducing the interaction between impurities and dislocations in crystal, Nelson predicted a hexatic-crystal-hexatic reentrant transition upon heating from $T = 0$ for any finite concentration of impurities below a critical value[172]. This view is contested by the analytical calculations and numerical simulations from Cha and Ferstig[173], who argued that the non-negligible effect of disordered configurations at low temperatures leads to the failure of fugacity expansion in the RG calculation by Nelson. The phase diagram plotted as a function of the impurity concentration and temperature suggests that in the bottom-left regime, i.e., low impurity concentration and low temperature, the system is stabilized in the crystal state, while it melts either at high impurity concentration or at high temperature.

It is heuristic to ask whether the impurity-driven melting (also referred to as "zero-temperature melting") and thermally driven melting obey the same mechanism. Impurities are mimicked by freezing a certain fraction of randomly selected particles in simulations[174-176]. The results indicate that these two pathways can be distinguished from only a single simulation snapshot[174]. There are no topological defects before melting at zero temperature, while free dislocations directly appear to release stress when the impurity concentration reaches a critical value, exhibiting a picture completely different from that of the KTHNY theory.

### 3.4. Numerical verifications of the KTHNY theory and the grain-boundary melting theory

Larger simulation scales allow direct quantitative verifications of the analytical theories described above. Quantitative comparison requires categorizing the identified topological defects into free disclinations, free dislocations, dislocation pairs, vacancies or those belonging to larger clusters. A detailed introduction to the related methodologies can be found in refs. [97, 177, 178]. Here we focus on three key aspects of the numerical examinations: (1) the fundamental physical picture of the KTHNY theory; (2) the critical exponents predicted by the RG analysis; (3) the theoretical consistency between the core energy and the melting scenario.

In the hard disk system, it has been found that almost no free dislocations appear in the crystal state, but there can be a small fraction of disclinations in the hexatic phase[178]. These disclinations are close to other larger defect structures, rather than free, and therefore do not destroy the bond-orientational order. Another simulation on the Yukawa system[135], where the potential is relatively soft and the KTHNY melting scenario is exhibited, suggested that free dislocations exist even in the crystal phase, but disappear after quenching. Whether this difference is universal across hard and soft particle systems remains unclear. On the one hand, these results are generally consistent with the basic assumptions of the KTHNY theory, and on the other hand, they reveal that real particle systems are more complex than the idealized theoretical assumptions. Moreover, the HNY RG analysis[11-13] predicts a critical behavior of dislocation/disclination density close to the critical point, which is $\rho_d = a\exp\{-b[T_c/(T_c-T)]^{\nu}\}$ where $\nu \simeq 0.37$ for the crystal-hexatic transition while $\nu \simeq 0.5$ for the hexatic-liquid transition[11-13]. A quantitative examination in the hard disk system[178] found that, for the crystal-hexatic transition, this scaling law fits the simulation data well when the exponent $\nu$ is fixed to the theoretical value. However, a four-parameter fit with the exponent $\nu$ as a free parameter can also yield good agreement, whereas the fitted exponent can be as large as 9, manifesting that the fitting is insensitive to this critical exponent. The authors also performed a fitting for the hexatic-liquid transition, but the transition here is first-order, and thus is not expected to exhibit the critical scaling behavior predicted for continuous phase transitions. Consequently, to date, there is no conclusive numerical evidence to either support or refute the theoretical critical exponents.

The defect structure close to the hexatic-liquid transition reveals interesting phenomena beyond the original picture of analytical theories. The hexatic phase is composed of many crystal-like clusters with different orientations. Defects appear at the boundaries of different clusters, which are similar to the grain boundaries. Performing simulations on hard and soft disks ( $\varepsilon(\sigma/r)^6$ )[178], the authors investigated both the discontinuous hexatic-liquid transition and the continuous one. In the KTHNY melting scenario, these grain boundaries percolate at the hexatic-liquid transition point. For a first-order transition, there is a phase coexistence region, in which the liquid clusters are rich in defects, and the transition to the liquid phase is then driven by the percolation of liquid clusters. The percolation threshold is close to the phase transition point, but whether percolation merely accompanies the melting process, or instead constitutes its underlying mechanism, remains unclear. The earlier simulation of the Yukawa system[135], which exhibits a continuous hexatic-liquid transition, also found that defects assemble into grain boundaries in the hexatic phase after quenching, and these grain boundaries percolate at the transition to the liquid phase.

Far less attention has been paid to the core energy, since the simulation results are sufficiently conclusive for determining the melting scenario. The calculation of the core energy, however, can tell us whether it is possible to predict the melting scenario based on the analytical theories. It has been suggested that the core energy can be calculated from the defect density as $P_d = \frac{16\sqrt{3}\pi^2}{K-8\pi} I_0(\frac{K}{8\pi})\exp(\frac{K}{8\pi})\exp(\frac{-2E_c}{kT})$ [111, 179]. The simulation result for the hard-disk system provides a core energy of 5.35$kT$[180], consistent with the continuous transition from crystal to hexatic. The underlying assumption of the above formula is that the defects are independent, resulting in an Arrhenius form of the defect density[70]. However, the statistics suffers from large errors at low temperatures, where only a small number of defects are present, and the influence of defect interactions is non-negligible at high temperatures. Another difficulty is that the core energy can only be calculated in the crystal state, where the Young's modulus remains finite, making it impossible to verify the theoretical boundary of the core energy in hexatic-liquid transitions[180].

## 4. 2D phase transitions in active matter systems

Active matter has recently become one of the hottest topics in soft matter physics. Various classical conclusions in statistical mechanics are validated and investigated in different active matter systems, advancing our understanding of the non-equilibrium statistical physics. Below we will discuss the interesting results in non-equilibrium XY models and non-equilibrium 2D crystals separately, highlighting the phenomena beyond the description of equilibrium statistical physics.

### 4.1. Non-equilibrium XY model

The pioneering study of the 2D Vicsek model[50] revealed a flocking state, where each "flying spin" points in the same direction. The emergence of long-range order seemingly violates the HMW theorem[3-5]. This unexpected observation was later phenomenologically explained by the Toner-Tu equation[181, 182]. Roughly speaking, although each spin in the Vicsek model still interacts with several nearest neighbors, the continuous self-propelled motion of the particles propagates orientational information across the entire system. The 2020 Onsager Prize was awarded to Vicsek, Toner, and Tu for their fundamental contributions to the field of active matter. Many subsequent studies[183-190] sought to clarify the phase diagram, phase transition order, metastability, and fragility of the Vicsek model. Moreover, various variants of the Vicsek model have been proposed[47], including those with a large self-propulsion velocity[191], topological rather than metric neighbor selection[192, 193], alternative noise types[194, 195], or bipolar alignment[196]. These models can exhibit first-order, second-order, or BKT phase transitions. A detailed discussion of these extended models is beyond the scope of this review.

The XY model is the simplest system that incorporates real dynamical evolution, as the spin orientational angle is a continuous variable, and its time evolution can be integrated step by step. Driven by the torque produced by neighbors, spins evolve into the equilibrium state of the XY model under Brownian dynamics. We now use an Ornstein-Uhlenbeck process to describe the active noise[51, 52], in which the time evolution of each spin can be written as $\dot{\theta}_i = -\frac{\partial H_{XY}}{\partial \theta_i} + \psi_i$ where the noise term $\psi_i$ evolves as $\tau\dot{\psi}_i = -\psi_i + \sqrt{T}\eta_i$ and $<\eta_i(t)\eta_j(t')>= 2\delta_{ij}\delta(t-t')$, or equivalently, has a finite correlation time of $\tau$, i.e., $<\psi_i(t)\psi_j(t')>= \frac{2T}{\tau}\delta_{ij}\exp(-|t-t'|/\tau)$. This persistent noise effectively captures the key effect of self-propulsion in active matter systems[45]. The correlation function can be derived by considering the continuous field of $\theta$. The above equations indicate that the spectrum of $\theta$ differs from that of the equilibrium XY model, leading to a correlation function decay with an exponent $\eta = \frac{\tau T}{2\pi k}$ that is proportional to $\tau$, as shown in Fig. 6(a). This relation indicates that the upper bound on the decay exponent set by the BKT theory[6-8] is no longer valid for such non-equilibrium systems.
The interaction in the XY model favors the alignment between neighboring spins. One of the most common mechanisms for the alignment in active matter systems comes from vision. Vision-based interactions naturally induce non-reciprocity, as a particle can only perceive other elements in front of it. Preliminary attempts at introducing a vision cone into the XY model found that the system exhibits true long-range order[54, 55]. However, this long-range ordered state has recently been argued to be metastable against defect formation[58], as its continuum description is equivalent to the constant-density Toner-Tu equation[186]. Simulations[58] manifested that, for a sufficiently large system, it exhibits a dynamic foam phase as depicted in Fig. 6(b), with a correlation length saturating as $T \to 0$, which verifies the absence of truly long-range order.
Compared with particle systems, the XY model is much simpler, and its continuum description is usually found to be effective. Moreover, the incorporation of real-time evolution enables straightforward Brownian dynamics simulations, avoiding the ambiguity arising from the misapplication of equilibrium simulation methods to non-equilibrium systems[197-199]. Therefore, the XY model provides an excellent platform for examining the effects arising from novel physical ingredients.

### 4.2. Morphologies and melting of non-equilibrium 2D crystals

The active Brownian particle (ABP) model[200, 201] is a simpler framework than the Vicsek model. The ABP system consists of self-propelled particles without alignment interactions. One of the most well-known results in the ABP system is the motility-induced phase separation (MIPS)[202], where liquid clusters emerge under solely volume-exclusion repulsion and the onset density of MIPS is far lower than the density of the gas-solid coexistence in passive hard-disk systems. Further addition of an effective alignment force arising from entropic effects[203] or an isotropic attraction[204] gives rise to a flocking state reminiscent of the Vicsek model. It is therefore instructive to ask whether the MIPS phase correlates with the specific nature of 2D melting.
In 2018, two groups of researchers[205, 206] focused on the phase diagram of the ABP system, utilizing the kinetic Monte Carlo (KMC) method and Brownian dynamics (BD) simulations, respectively. The KMC simulation results[205] suggested that the MIPS regime has no correlations with 2D melting, and was merely a small region within the fluid phase analogous to conventional liquid-gas coexistence; by contrast, BD simulation results[206] indicated that the MIPS regime overlapped with the hexatic-liquid transition at high densities and high activities. The parameters used in these two works are different, making it difficult to assess the consistency of their results. In these studies, phase boundaries were determined based on the upper bounds on the correlation function decay exponents predicted by the KTHNY theory: 1/3 for the translational one while 1/4 for the bond-orientational one. Shi et al.[207] simulated an ABP model with XY-like alignment interactions, and found that the active crystal could sustain a large deformation, leading to a larger decay exponent of the translational correlation function, with an example illustrated in Fig. 6(c). The underlying mechanism was attributed to the persistent noise, analogous to that of the active Ornstein-Uhlenbeck XY model discussed earlier. Based on the linear elastic theory, they successfully predicted the decay exponent at the crystal-hexatic transition, which however, did not show a universal value. It remains unclear whether these findings hold for original ABP systems, i.e., whether the decay exponent can exceed the 1/3 upper bound in the absence of alignment interactions.

Digregorio et al.[178] systematically investigated the melting scenario of both the passive and active hard disk systems. Although the melting scenario changes from the hard-disk-like behavior to the KTHNY theory at high activity, they suggested that the critical behavior and the percolation transition near the hexatic-liquid transition were largely independent of activity. It is important to note that they adopted phase boundaries determined in their earlier work[206], which employed criteria based on equilibrium statistical mechanics. Therefore, these results may just suggest that the decay behavior of correlation functions is still determined by topological defects in active systems. Notably, this conclusion has also been confirmed in a completely different system consisting of charged colloids and active spinners[208]. It is therefore natural to ask: if the statistical mechanics of topological defects remains unchanged, what is the origin of the difference between active and passive melting? One candidate for this difference is that the temperature is generally not a well-defined thermodynamic quantity in active matter systems[43, 48, 209, 210], which may lead to a more complex RG analysis. Moreover, the lack of a Hamiltonian in most active matter systems necessitates the dynamic RG approach instead of an equilibrium one[211]. This fundamental open question represents a key direction for future research.

Another model system designed for the simulation of absorbing phase transitions[212, 213] has also been employed for the study of 2D active crystals. In this system, Galliano et al.[214] observed the truly long-range translational order. They took as evidence the fact that the MSD did not diverge with the system size (see Fig. 6(d)) and the spectrum of displacement lacked the $q^{-2}$ scaling at small $q$, which are the defining features of the long-range order. Roughly speaking, this phenomenon still originates from the breakdown of the equipartition theorem in non-equilibrium systems. In addition, Voronoi models with self-propulsion[215], Ornstein-Uhlenbeck noise[216], and cell division[217] have been simulated. Specifically, by introducing a pulsating activity, Li et al.[218] observed rich phenomena including the BKT phase transition regarding the pulsating phase, giant fluctuations, and hyperuniformity. All these results are well captured by their phenomenological hydrodynamic equations.

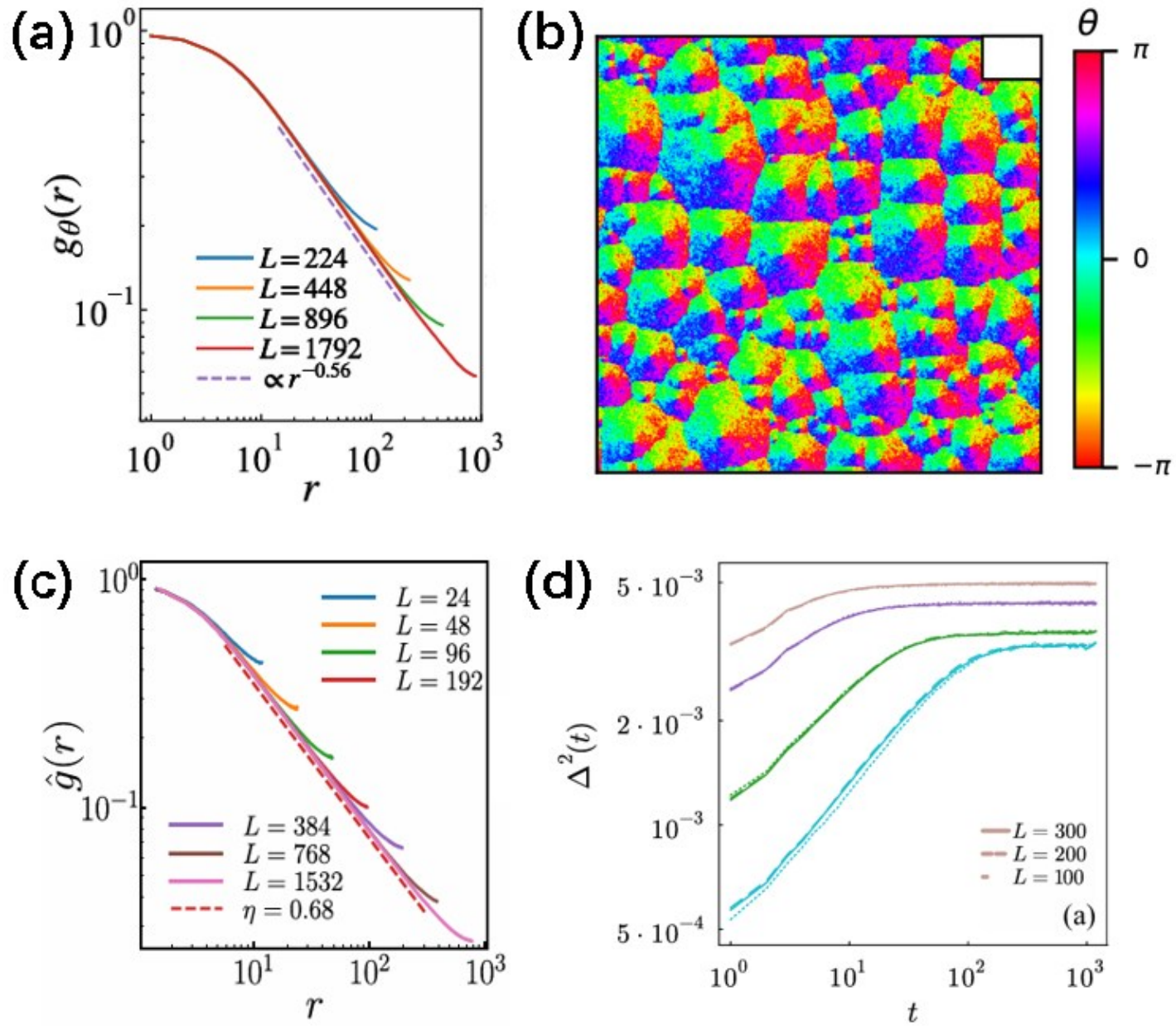

Fig. 6 Phase behaviors of the non-equilibrium 2D systems. (a) The spin-spin correlation function in the XY model with a persistent noise[51]. The decay exponent can be much larger than the upper bound predicted by equilibrium statistical mechanics (0.25). (b) Simulation snapshot of the XY model with a vision cone, forming a dynamic foam phase[58]. (c) The translational correlation function in the 2D crystal phase of ABP with alignment interaction[207]. The decay exponent exceeds the upper bound predicted by equilibrium statistical mechanics (1/3). (d) The MSD curve for systems with different sizes in a system designed for absorbing transition[214]. The plateau of different system sizes in each case remains unchanged, manifesting a truly long-range order.

## 5. Summary and outlook

It has been 60 years since the formulation of the landmark HMW theorem, which has inspired pioneering ideas of topological defects, topological phases, and infinite-order phase transitions, and exerted a far-reaching influence on statistical physics, condensed matter physics, soft matter physics, and even biophysics. The BKT transition is found to be general for continuous symmetries in 2D. Notably, its application to 2D melting reveals novel phenomena, which are distinct from the simple first-order solid-liquid phase transition in 3D. Through detailed HNY RG analysis, the so-called KTHNY theory reveals two BKT transitions separated by a unique hexatic phase. However, unlike the well-established BKT transition in 2D spin systems, the predictions of the KTHNY theory are not always valid for 2D melting. The grain-boundary theory and the simulation of the hard-disk system suggest two alternative melting scenarios. Most analytical theories have not been rigorously tested even to date, while a wealth of phenomena have been observed that lie beyond the existing analytical frameworks. Leveraging numerical simulation results, qualitative predictions of the melting scenario can be made, but quantitative predictions for a specific system remain challenging. The Vicsek model is a celebrated example of the seeming violation of the HMW theorem in non-equilibrium systems, sparking the interests in active matter physics, in which various systems can exhibit very different phase behaviors. Although some of these phenomena are well understood – for example, the effect of self-propulsion can usually be described by persistent noise – most observed behaviors still lack a general theoretical framework.

We would like to point out several possible future directions for this area, regarding both classical 2D melting and the phase behavior in 2D non-equilibrium systems. After the extensive simulations on different model systems, the most important problem in classical 2D melting is prediction, including the appearance of the hexatic phase, the order of a phase transition, and the melting point. As we have discussed in Sec. 2, a quantitative prediction based on analytical theories faces too many problems. Therefore, we provide another general route towards qualitative predictions based on the original idea of statistical mechanics. Recall that, back to the ensemble theory provided by Gibbs, for a given Hamiltonian $H$ as a function of all the momenta and coordinates, the distribution function in the phase space is just $\rho(\vec{r}^N, \vec{p}^N) \sim \exp(-\beta H)$ and all the thermal quantities can be calculated by averaging over this distribution function. This fundamental idea reminds us that we may concentrate on monomer properties and build up the connection between these properties and thermodynamics. For a given particle, its ground state depends on the interaction (enthalpic contribution) and its shape (entropic contribution); both elements determine the local structure. The local structure further determines the thermodynamic behaviors. As we have discussed in Sec. 3, this route can approach a reasonable prediction on the hexatic phase and can provide important indications for the phase transition order. Further investigations of those phenomenological rules should be performed in more complex lattices, such as honeycomb and Kagome. Another important factor is kinetics/dynamics, which also relies on the local structure and can significantly affect the thermodynamics. Non-Fickian diffusion and spatial heterogeneity have already been observed in 2D crystals[219-222], and the hopping process has been investigated in the hexatic phase[223]. Less attention is paid to the connection between these dynamic features and the melting process, but we believe the dynamics differs fundamentally between different kinds of melting scenarios. A schematic summary of this route is illustrated in Fig. 7.

The prediction of the melting point is rather challenging. Analytical theories can be applied to the crystal-hexatic transition, since Young's modulus, which has a well-defined microscopic description, approaches a universal value at the phase transition point. Although there exist some difficulties as we mentioned in Sec. 2, it is possible to provide a reasonable estimation over a certain range of parameters. However, this analytical approach cannot be extended to predict the hexatic-liquid transition. A good candidate is the use of neural networks, which have been shown to be powerful for learning the mapping between input features and target outputs from training data. On the one hand, it is of limited practical value to predict the phase transition point of model systems, since they do not directly correspond to any realistic systems; on the other hand, there are few data for realistic 2D systems. Consequently, we should use the data from model systems as input, and learn a general as well as quantitative mapping to the melting point. There are, however, several fundamental difficulties: (i)

There are relatively few available data compared to datasets for proteins[224, 225] or crystals[226, 227]; (ii) Finite-size simulations need to be enhanced by utilizing some recently developed methods[228-230] in order to reach quantitatively accurate results; (iii) Most of our physical understanding on this problem is intuition rather than formulae, and this intuition-based knowledge is difficult to incorporate into neural networks.

The hexatic phase, located between order and disorder, has been observed in many realistic systems, such as confined water[31, 32], quantum systems[21, 26, 30], and different kinds of colloids[34-36]. It is worthwhile to explore the potential applications of the hexatic phase. Moreover, a hexatic-tetratic transition has been reported in a soft-core system[176]. Both solid-solid[231-240] and liquid-liquid transitions[241-244] have been extensively studied and known to be essential for applications; the hexatic-tetratic transition lies between these two classes of phase transitions, possibly providing new fundamental insights into the kinetics of phase transitions.

Turning to active matter systems, we highlight several promising directions for future discovery. The non-reciprocity arising from the vision cone has been well studied, but another major source of non-reciprocity – that arising from hydrodynamic interactions[245-247], which is ubiquitous in bacterial suspensions[248] – remains poorly understood[249]. Research on active matter systems teaches us that the relationship between topological defects and the correlation function is robust; the marked differences between active and passive melting should therefore be attributed to either the excitation of topological defects or their intrinsic motion. Topological defects have also been widely defined in amorphous solids, typically via the associated stress or strain fields[250, 251]. Whether topological defects in these systems still follow the statistical mechanics of the Coulomb gas model is worthy of investigation. Exploring these fundamental topics will undoubtedly not only advance our understanding of non-equilibrium physics, but also deepen our insight into the intrinsic relation between geometry and physics.

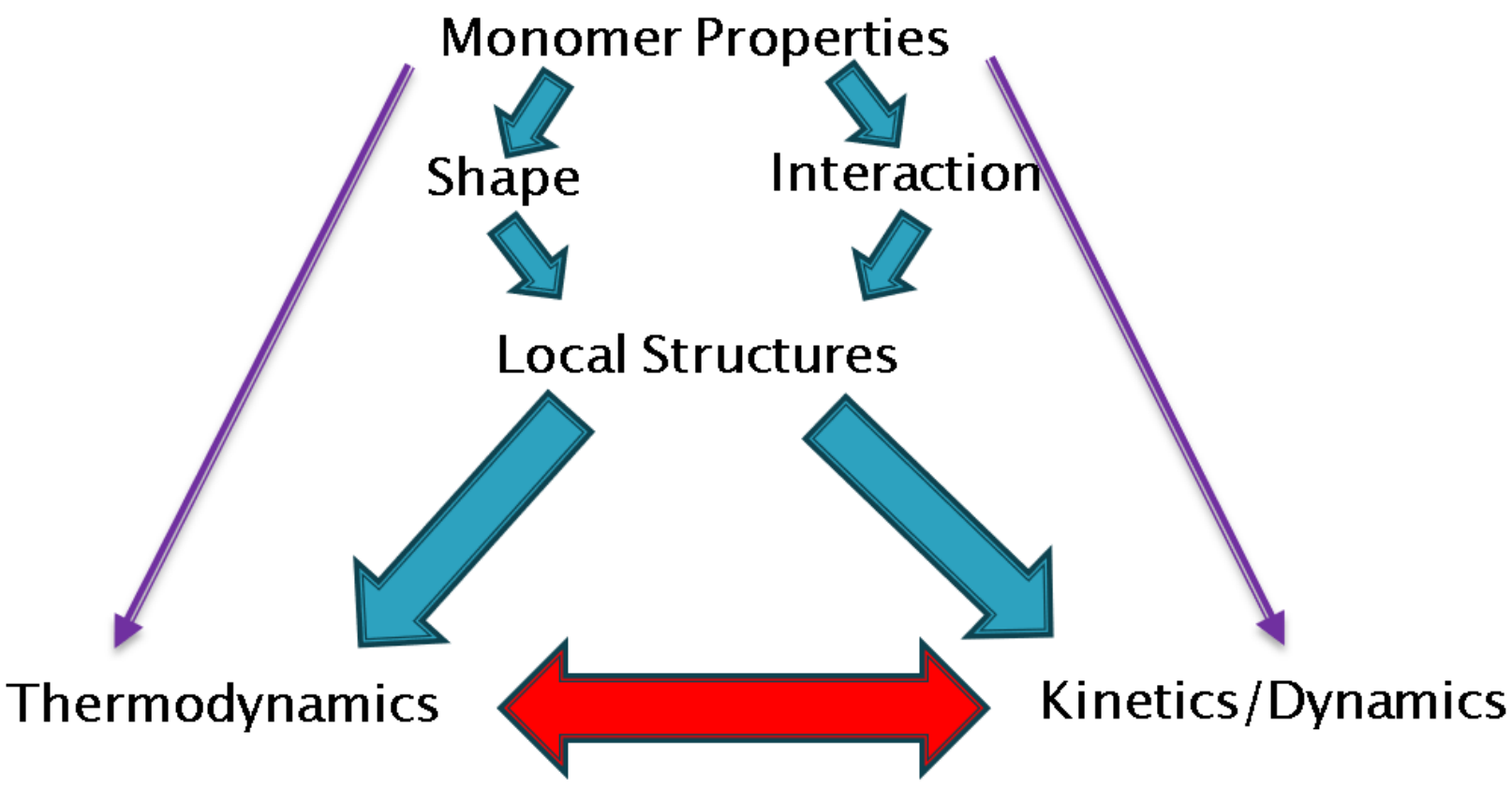

Fig. 7 Theoretical route towards the prediction of 2D melting scenarios based on monomer properties of shape and interaction, which directly determine the equilibrium local structure and are known to be crucial to both thermodynamics and kinetics/dynamics. The connection between thermodynamics and kinetics/dynamics is still severely under-explored.

**Acknowledgment**
This work was supported by the National Natural Science Foundation of China (Nos. 12574236, 12447101).